\documentclass[runningheads]{llncs}
\usepackage{graphicx}
\usepackage[firstpage]{draftwatermark}
\SetWatermarkText{\hspace*{11cm}\raisebox{14.8cm}{\includegraphics{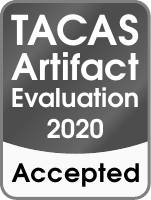}}}
\SetWatermarkAngle{0}

\usepackage{amsmath}
\usepackage{times}
\usepackage{cite}
\usepackage{fancyvrb}
\usepackage{bibnames}
\usepackage{graphicx}
\usepackage[pdftex,dvipsnames]{xcolor}  
\usepackage{amsmath,amssymb}
\usepackage{listings}
\usepackage{multirow}
\usepackage{longtable}
\usepackage{arydshln}
\usepackage{enumitem}
\usepackage{amsfonts}
\usepackage{multicol}
\usepackage{algorithm}
\usepackage[noend]{algpseudocode}
\usepackage{xargs}
\usepackage[colorinlistoftodos,prependcaption,textsize=tiny]{todonotes}
\usepackage{wrapfig}

\usepackage{floatflt}
\usepackage{mdframed,framed}

\makeatletter
\def\BState{\State\hskip-\ALG@thistlm}
\makeatother

\graphicspath{ {./figures/} }

\newcommandx{\unsure}[2][1=]{\todo[linecolor=red,backgroundcolor=red!25,bordercolor=red,#1]{#2}}
\newcommandx{\change}[2][1=]{\todo[linecolor=blue,backgroundcolor=blue!25,bordercolor=blue,#1]{#2}}
\newcommandx{\info}[2][1=]{\todo[linecolor=OliveGreen,backgroundcolor=OliveGreen!25,bordercolor=OliveGreen,#1]{#2}}
\newcommandx{\improvement}[2][1=]{\todo[linecolor=Plum,backgroundcolor=Plum!25,bordercolor=Plum,#1]{#2}}
\newcommandx{\thiswillnotshow}[2][1=]{\todo[disable,#1]{#2}}

\newcommand{\ProbModel}{Prob-solvable}
\newcommand{\Mora}{\textsc{Mora}}

\makeatletter
\RequirePackage[bookmarks,unicode,colorlinks=true]{hyperref}%
   \def\@citecolor{blue}%
   \def\@urlcolor{blue}%
   \def\@linkcolor{blue}%

\def\orcidID#1{\smash{\href{http://orcid.org/#1}{\protect\raisebox{-1.25pt}{\protect\includegraphics{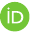}}}}}
\makeatother

\begin{document}

\title{\textsc{Mora} - Automatic Generation of Moment-Based Invariants}
\titlerunning{ }
\authorrunning{ }
\author{ }
\author{Ezio Bartocci\inst{1}\orcidID{0000-0002-8004-6601} \and
Laura Kov{\'{a}}cs\inst{1,2}\orcidID{0000-0001-8263-8689} \and
Miroslav Stankovi\v{c}\inst{1}\orcidID{0000-0001-5978-7475}}
\institute{TU Wien, Vienna, Austria \and Chalmers, Gothenburg, Sweden}

\maketitle   
\let\thefootnote\relax\footnotetext{
This research was supported by 
the ERC Starting Grant 2014 SYMCAR 639270, 
the Wallenberg Academy Fellowship 2014 TheProSE, 
and the Austrian FWF project W1255-N23.}

\begin{abstract}
We introduce \textsc{Mora}, an automated tool for generating
 invariants of probabilistic programs. Inputs to \textsc{Mora} are
 so-called  \ProbModel~loops, that is probabilistic programs with
 polynomial assignments over random variables and parametrized
 distributions. Combining methods from symbolic computation and
 statistics, 
 \textsc{Mora} computes invariant properties over higher-order moments
 of loop variables, expressing, for example, statistical properties, such
 as expected values and variances, over the value distribution of loop
 variables.
\end{abstract}
\vspace{-5pt}
\section{Introduction}
\label{sec:intro}
\vspace{-5pt}

Probabilistic programs (PPs) are becoming more and more commonplace.
Originally employed in randomized algorithms
and cryptographic/privacy protocols, now gaining momentum due to the several emerging applications in the areas
of machine learning and AI~\cite{Ghahramani15}.
By introducing randomness into the program, program variables can no
longer be treated as having single values; we must think about them as distributions. 
Dealing with distributions is much more challenging and some simplifications are required. 
Existing approaches, see e.g.~\cite{McIverM05,Katoen2010,Chakarov2014,Barthe2016,Kura19}, 
usually take into consideration only expected values or upper and lower bounds over 
program variables, or rely on user guidance for providing templates and hints. 

One of the main challenges in analyzing PPs and computing their
higher-order moments comes with the presence of loops and the burden
of computing so-called \emph{quantitative
  invariants}~\cite{Katoen2010}. Quantitative invariants are
properties that are true before and after each loop
iteration and are crucial for analyzing the
behavior of PP loops.

In this paper, we introduce the \textsc{Mora} tool for computing
quantitative invariants of a class of PPs, called {\it \ProbModel{} loops}~\cite{Bartocci2019},
with random assignments, parametrized distributions, and polynomial
probabilistic updates.
   Our implementation is available at:
\begin{center}\vspace{-5pt}
\url{https://github.com/miroslav21/mora},
 \end{center}\vspace{-5pt}
   and successfully evaluated on
a number of challenging examples.
Unlike other existing approaches, 
e.g.~\cite{Katoen2010, Chakarov2014,Barthe2016,Kura19}, \textsc{Mora} 
computes non-linear invariants in a fully
automatic way, without relying on user-provided
templates/hints.  The proposed automatic approach 
can handle an arbitrary number of loop iterations and also infinite loops.
On the contrary, tools like PSI~\cite{GehrMV16} support only the 
automatic analysis of probabilistic programs with a specified 
number of loop iterations.

Moreover,  the invariants inferred by \textsc{Mora}
are not restricted to expected values but are quantitative
invariants over the higher-order moments of
program variables. We refer to such invariants as {\it moment-based
invariants}~\cite{Bartocci2019}.  To the best of our knowledge, no other approach
can so far automatically compute higher-order moments of PPs, not even for the
restricted yet expressive enough class of  \ProbModel{} loop.

\begin{figure}[t]
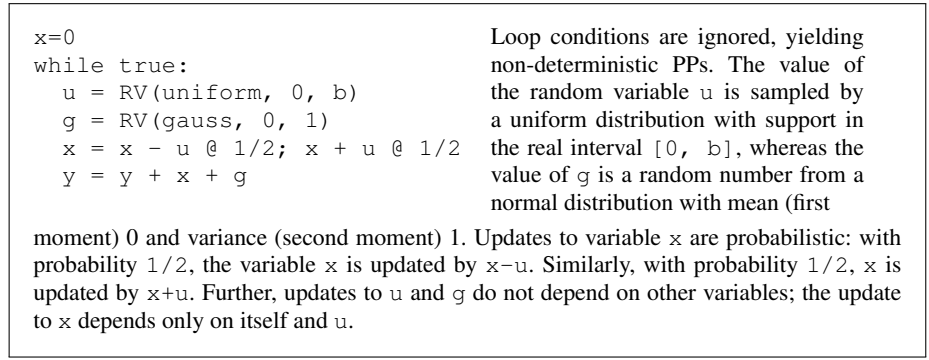

  
    \begin{framed}
    \begin{minipage}[t]{0.52\textwidth}
\begin{verbatim}
x=0
while true:
  u = RV(uniform, 0, b)
  g = RV(gauss, 0, 1)
  x = x - u @ 1/2; x + u @ 1/2
  y = y + x + g
\end{verbatim}
    \end{minipage}
    \begin{minipage}[t]{0.43\textwidth}
      {\small Loop conditions are ignored, yielding non-deterministic PPs. 
      The value of the random variable {\tt u} is sampled by a uniform
      distribution with support in the real interval {\tt [0, b]},
      whereas the value of {\tt g} is a random number from a normal
      distribution with mean (first  }
  \end{minipage}
  \vskip.5em
  {\small moment) 0 and variance (second
      moment) 1.  Updates to variable {\tt x}  are probabilistic:
      with probability {\tt 1/2}, the variable {\tt x} is updated
      by 
      {\tt x-u}. Similarly, with probability {\tt 1/2}, {\tt x} is
      updated by {\tt x+u}. Further, updates to {\tt u} and {\tt g} do
    not depend on other variables; the update to {\tt x} depends only
     on itself and {\tt u}.}
      \end{framed}
      \vspace{-15pt}
      \caption{An illustrative example\label{fig:running} of a \ProbModel{}
  loop. }
  \vspace{-5pt}
\end{figure}

The purpose of this paper is to describe what \textsc{Mora} can do and
how it can be used. The paper is intended as a tool demonstration and guide for potential
users of \Mora. 
We focus on  the usage and implementation aspects of
\textsc{Mora}. For details on theoretical foundations and  algorithmic
aspects of \textsc{Mora} for computing moment-based invariants, we
refer to~\cite{Bartocci2019}. We note however that, when compared
  to the experimental setup of~\cite{Bartocci2019},  \textsc{Mora}
comes with a completely new design, fully
implemented in {\tt python} and supporting an easy installation and use by
even non-experts in PPs.

\vspace{-5pt}
\section{\Mora -- Programming Model}\label{sec:pgm}
\vspace{-5pt}
Input programs  to \Mora{} are PP loops that are
\ProbModel~\cite{Bartocci2019}. In Figure~\ref{fig:running}, we give
an example of a \ProbModel{} loop and use this example as a running
example to guide the potential users of \Mora{} in the rest of this
paper. 

In a nutshell,  the probabilistic
assignments of \ProbModel{} loops 
involve 
(i) variable values drawn from random distributions, such as uniform or
normal distributions, and (ii) random variable updates. In
the sequel, we write {\tt RV} to refer to a random
variable. Input programs to \Mora{} thus satisfy the
following two properties:

\noindent (1) Input programs to \Mora{} are  PPs 
generated from the grammar in Figure~\ref{fig:grammar}.

\begin{figure}
\begin{framed}
  \begin{minipage}[t]{0.9\textwidth}
\center   \underline{Grammar defining PP inputs to \Mora{}}
    \lstset{basicstyle=\scriptsize, literate={->}{$\rightarrow$}{2}}
\begin{lstlisting}
PROGRAM -> INIT_ASSIGNS "while true:" RV_ASSIGNS UPD_ASSIGNS

INIT_ASSIGNS -> INIT_ASSIGN | INIT_ASSIGN INIT_ASSIGNS
RV_ASSIGNS -> RV_ASSIGN | RV_ASSIGN RV_ASSIGNS
UPD_ASSIGNS -> UPD_ASSIGN | UPD_ASSIGN UPD_ASSIGNS

INIT_ASSIGN -> VAR " = " INIT_EXPR
RV_ASSIGN -> VAR " = " RV_EXPR
UPD_ASSIGN -> VAR " = " UPD_BRANCHES

UPD_BRANCHES -> UPD_BRANCH | UPD_BRANCH UPD_BRANCHES
UPD_BRANCH -> UPD_EXPR "@" UPD_PROB 
UPD_PROB -> SIMP_EXPR

INIT_EXPR -> RV_EXPR |  SIMP_EXPR
RV_EXPR ->  "RV(uniform, " SIMP_EXPR ", " SIMP_EXPR ")"
		| "RV(gauss, " SIMP_EXPR ", " SIMP_EXPR ")"
UPD_EXPR -> UPD_EXPR OP UPD_EXPR | VAR | ATOM
SIMP_EXPR -> SIMP_EXPR OP SIMP_EXPR | ATOM 

ATOM -> NUM | PARAMETER
OP -> [*+-]
VAR -> [a-zA-Z][a-zA-Z0-9]*
PARAMETER -> [a-zA-Z][a-zA-Z0-9]*
NUM -> [-]?[0-9]+[.]?[0-9]*([\/][1-9][0-9]*)?
\end{lstlisting}
  \end{minipage}
  \end{framed}
  \vspace{-15pt}
  \caption{}\label{fig:grammar}
  \vspace{-5pt}
  \end{figure}
  
\noindent (2) 
In addition to the grammar of Figure~\ref{fig:grammar}, \Mora{} requires  its PP input to be
\ProbModel, imposing further restrictions as follows: 
\begin{itemize}
	\item PP loop variables are different from each other and from parameters;
	\item probabilities used within  a variable update sum
          up to 1;
	\item updated variables  depend on themselves linearly and may
          depend polynomially only on other variables that have been previously updated.
\end{itemize}

Note that Figure~\ref{fig:running} satisfies all constraints above,
and thus is \ProbModel.

\vspace{-5pt}
\section{\Mora -- Usage}
\vspace{-5pt}
We describe the easiest way \Mora{} can be used to generate
moment-based invariants:
\begin{itemize}
  \item Save a
\ProbModel{} loop to a file, for example save 
Figure~\ref{fig:running} in the file {\tt running}
\item In the main \Mora{} folder invoke {\tt python} with {\tt
    python3.7} and execute:  \vspace*{-.5em}

  \begin{center}{\tt from mora.mora import mora}\vskip-.5em\end{center}
  
\item Run \Mora{} using the command: \vspace*{-.5em}
  \end{itemize}
\begin{center}
  {\tt mora("running", goal=GOAL)},
\end{center}
where {\tt GOAL} can be (i)
a specific natural number $k\geq 1$, in
which case \Mora{} computes the $k$th moments of all variables from
{\tt running}; (ii) a specific moment of one loop variable
of {\tt running} (e.g. {\tt "x\^{}2"} specifying the second moment of a
variable {\tt x} of Figure~\ref{fig:running}); or (iii) a list containing the 
goals as just specified. One can specify finitely many goals as inputs to
\Mora{}; yet, at least one goal is required. 
For example, by running {\tt mora("running", [1, "x\^{}2", "x\^{}3"])},
\Mora{} computes the expected values (first moments, i.e. 1) of all
variables from Figure~\ref{fig:running}, as well as the second and
third moments of variable {\tt x} of Figure~\ref{fig:running}
(specified by {\tt x\^{}2} and {\tt x\^{}3}, respectively).

\Mora{} is completely automatic.
That is, once an execution of \Mora{} is started on a given
\ProbModel{} loop and input goals, \Mora{} outputs the   
higher-order moments, and thus moment-based invariants, of its loop
w.r.t. the specified input goals.  
To this end, \Mora{} computes the
expected values of all monomials over loop variables, on which 
one of the goals from {\tt Goal} depends. 
In general, computing the {\tt k}th
moment requires computing the expected values of all monomial
expressions over loop variables, such that the 
total 
degree of the monomials is less or equal than $k$ -- see~\cite{Bartocci2019} for more details. 

In the rest of the paper, we will illustrate the main steps of
\Mora{}, by considering Figure~\ref{fig:running} as its input loop 
and {\tt [1, 2]} as its list of input goals. With such an
input goal, \Mora{} is set to compute the first 
and second moments of each variable of
Figure~\ref{fig:running}. Note, that even if {\tt 1} was omitted
from the aforementioned input goal, \Mora{} would still need to compute some 
of the first moments of the variables, as they are required for computing 
the second-order moments. In the sequel, we show-case the \Mora{} 
behaviour for: 
\begin{equation}\label{eq:running}
  \texttt{\tt mora("running", [1, 2])}.
 \end{equation}

\vspace{-5pt}
\section{\Mora -- Tool Overview} 
\label{sec:tool}
\begin{figure}[t]
\includegraphics[width=1\textwidth]{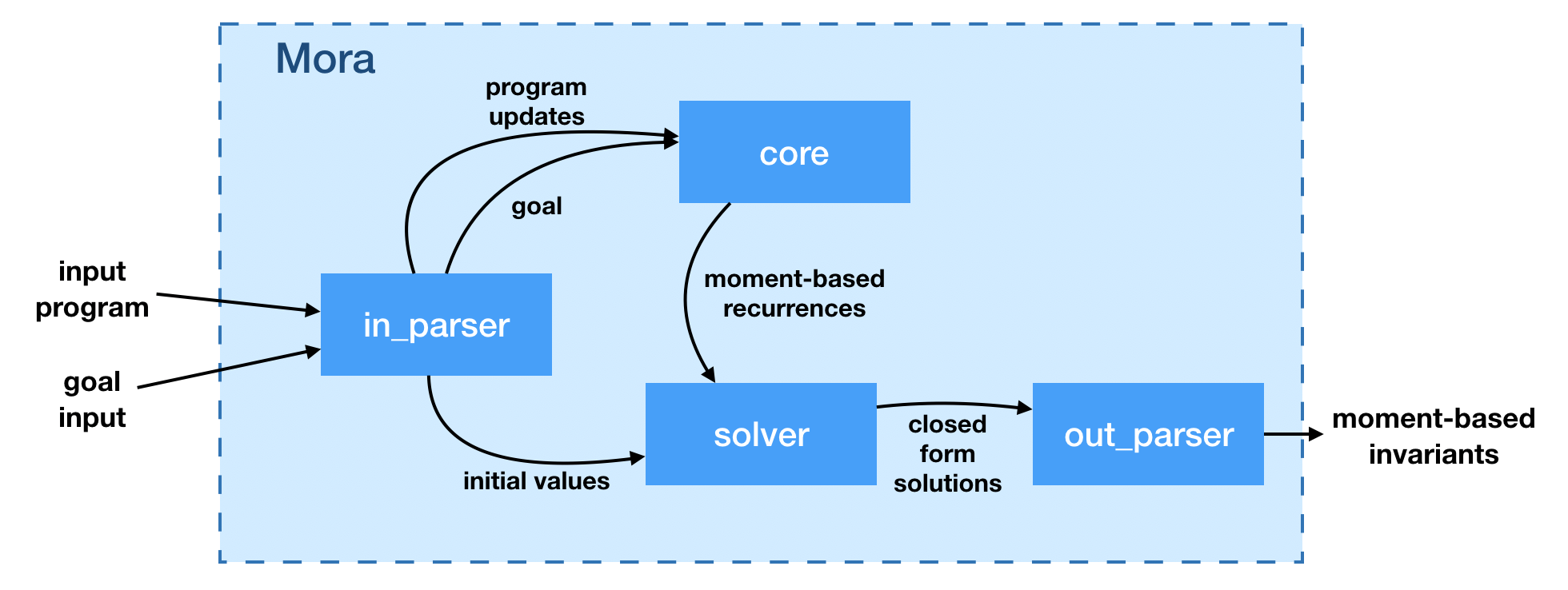}
\centering
\vspace{-15pt}
\caption{\textsc{Mora} workflow diagram.\label{fig:flow}}
\vspace{-5pt}
\end{figure}

\vspace{-5pt}
We first give details on our
implementation. We then present the overall  workflow of \textsc{Mora}  
in
Figure~\ref{fig:flow}, based on which we overview the main components
of our tool.

\noindent\paragraph{\textbf{Overall Implementation.}} \textsc{Mora} is
implemented in {\tt python3}, requiring  {\tt python} version of at
least 3.7. \Mora{}  relies on the {\tt diofant}  and {\tt scipy}
libraries: (i) the {\tt python} library {\tt diofant} is used in
\Mora{} for symbolic mathematical computations and recurrence solving;
(ii) the {\tt scipy} library, and in particular its statistics module
{\tt scipy.stats}, is used in \Mora{} to handle probability
distributions and statistical functions, as well as to simplify and
compute  expressions involving probability distributions and 
initial values of variables.  Altogether, our implementation
comprises of around 350 lines of code.

\noindent\paragraph{\textbf{\Mora{} -- Parser.}}
\Mora{} first checks whether a given input program is \ProbModel{}, by
checking the requirements of Section~\ref{sec:pgm}. If the input
program is not 
\ProbModel{}, an error
is reported, and the execution of \Mora{} stops. Otherwise, within
its parser module, \Mora{} extracts initial values from its input
loop, rewrites loop updates into equations over expected values of
monomial expressions over loop variables, and processes the list of its input
goals to identify which higher-order moments need to be computed.

For our demo execution~\eqref{eq:running}, \Mora{} extracts  the
initial value {\tt x(0)=0}, where {\tt x(0)} denotes the initial value of {\tt
    x} before the loop. Using the input goals specified
  in~\eqref{eq:running}, \Mora{} is set to compute the expected
  values of {\tt \{u, g, x, y, u\^{}2, g\^{}2, x\^{}2, y\^{}2}
  characterizing the first and second moments of all loop variables of
  Figure~\ref{fig:running}. Further, the loop
  updates of  Figure~\ref{fig:running} are rewritten by \Mora{} into
  equations over expected values, as follows:
  \vspace{-5pt}
  \begin{equation}\label{eq:expValues}
    {\Bigg\{}
    \begin{array}{l}
           E[x^k(n+1)] = E[1/2\cdot(x(n)-u(n+1))^k +
           1/2\cdot(x(n)+u(n+1))^k]\\
           E[y^k(n+1)] = E[(y(n) + x(n+1) + g(n+1))^k]
         \end{array},
       \end{equation}
       where $n\geq 0$ is the loop counter of
       Figure~\ref{fig:running}, $x(n)$ denotes the value of {\tt x}
       at the $n$th loop iteration, and $E[expr]$ is the expected
       value of an expression $expr$. 

\noindent\paragraph{\textbf{\Mora{} -- Core.}} After rewriting 
probabilistic loop updates into equations over expected values,
\Mora{} rewrites these equations into non-probabilistic recurrences
over so-called E-variables, with the loop counter $n$ being the
recurrence index.
E-variables are simply variables created
from monomials over original variables. Thanks to the restrictions
defining PPs to be \ProbModel{}, the resulting recurrences are linear
recurrences with constant coefficients, that is C-finite recurrences,
whose closed forms can always be computed~\cite{Kauers11}.  \Mora{}
solves these recurrences by calling its {\it Solver} module. 

Using the equations~\eqref{eq:expValues} over expected values,  the
non-probabilistic recurrences of Figure~\ref{eq:running} generated by
\Mora{} are as follows, using the \Mora{} synthax:
\vspace{-5pt}
\begin{equation}\label{eq:MoraRec}
  \begin{array}{l}
    {\tt y  =  x + y}\\
    {\tt g**2  =  1}\\
    {\tt x  =  x}\\
    {\tt u  =  b/2}\\
    {\tt x**2  =  b**2/3 + x**2}\\
    {\tt u**2  =  b**2/3}\\
    {\tt y**2  =  b**2/3 + x**2 + 2*x*y + y**2 + 1}\\
    {\tt g  =  0}\\
    {\tt x*y  =  b**2/3 + x**2 + x*y}
   \end{array}
  \end{equation}
The left-hand sides of these equations represent values of E-variables
at iteration $n+1$, while monomials over original variables on the
right-hand side represent E-variables at iteration $n$. For example,
the first equation of~\eqref{eq:MoraRec} stands for $E[y(n+1)] =
E[x(n)] + E[y(n)]$. On the other hand, the fourth equation
of~\eqref{eq:MoraRec} represents $E[x(n+1)^2]=\frac{b^2}{3}+E[x(n)^2]$,
as $b$ is a constant parameter and {\tt x**k} in {\tt python} denotes the $k$th power of $x$. 

\noindent\paragraph{\bf Solver.}
In this module, \Mora{} extracts and solves recurrences from the
non-probabilistic equations over E-variables computed by its {\it
  Core} module. By
exploiting the structure of \ProbModel{} programs, \Mora{} also
optimizes the order in which recurrences are solved, e.g. independent
recurrences are solved first. Partial solutions can be used to reduce 
the complexity of the latter recurrences. \Mora{} then uses the {\tt diofant} 
library to handle and solve single recurrences. 

For Figure~\ref{eq:running}, using the E-variable equations
of~\eqref{eq:MoraRec}, the following closed form solutions are
computed by \Mora{}: 
\vspace{-5pt}
\begin{equation}\label{eq:MoraCF}
  \begin{array}{l}
E[u^2] = \frac{b^{2}}{3}\\
E[x^1] = 0\\
E[y^1] = y(0)\\
E[x^2] = \frac{b^{2} n}{3}\\
E[u^1] = \frac{b}{2}\\
E[y^1x^1] = \frac{b^{2} n}{6} \left(n + 1\right)\\
E[y^2] = \frac{n}{18} \left(2 b^{2} n^{2} + 3 b^{2} n + b^{2} + 18\right) + y(0)^{2}\\
E[g^1] = 0\\
    E[g^2] = 1\\
    \end{array}
\end{equation}
with $y(0)$ standing for the  initial value of $y$ (treated as
a parameter, since not specified).
]

\noindent\paragraph{\textbf{\Mora{} -- Out\_Parser.}}
\Mora's output consists of basic information about the 
program and the goal, moment-based invariants computed, 
and computation time. By default, the \Mora{} output is shown only on the 
screen. However, an optional argument can specify 
if an output file should be created. Two possible values 
for {\tt output\_format} are (i) {\tt "txt"}, producing a simple 
human-readable file, and (ii) {\tt "tex"}, producing a file with 
invariants in \LaTeX\ format (as given in~\eqref{eq:MoraCF} above).

\vspace{-7pt}
\section{Evaluation}\vspace{-5pt}
\label{sec:evaluation}

A proof-of-concept implementation, together with initial experiments,
were already given in our work on generating
moment-based invariants~\cite{Bartocci2019}. \Mora{} comes however
with a new design and re-implementation of~\cite{Bartocci2019},
significantly improving the experimental setting and evaluations
of~\cite{Bartocci2019}.
Table~\ref{table:compare} compares \Mora{} against the experiments
of~\cite{Bartocci2019}, on a subset of \ProbModel{} loops
from~\cite{Bartocci2019}, evidencing that \Mora{} is faster than our
initial proof-of-concept implementation. 
This is due to the 
following reasons:
\begin{wraptable}{r}{0.45\textwidth}
\vspace{-20pt}
\scriptsize
\begin{center}
\begin{tabular}{ | l | c | p{11mm} | p{11mm} | }
\hline
\textbf{Program}	& \textbf{Moment}	& \textbf{Runtime PoC ($s$)}	& \textbf{Runtime \Mora{} ($s$)}    \\ 
\hline\hline
\multirow{3}{*}
 {\textsc{sum\_rnd\_series}} 	& 1 	& 0.31 		& 0.22	\\
 \cline{2-4}							& 2 	& 2.89 		& 0.93	\\
 \cline{2-4}							& 3 	& 17.7 		& 2.47 	\\\hline
\multirow{3}{*}
 {\textsc{stutteringA}} 			& 1 	& 0.44 		& 0.25 	\\
 \cline{2-4}							& 2 	& 2.20 		& 1.07 	\\
 \cline{2-4}							& 3 	& 8.48 		& 3.35  	\\\hline
\multirow{3}{*}
 {\textsc{stutteringC}} 			& 1 	& 1.80 		& 0.66 	\\
 \cline{2-4}							& 2 	& 72.5 		& 12.2 	\\
 \cline{2-4}							& 3 	& 2144 		& 73.9  	\\\hline
\multirow{3}{*}
 {\textsc{Square}} 					& 1 	& 0.38 		& 0.22 	\\
 \cline{2-4}							& 2 	& 2.46 		& 0.73 	\\
 \cline{2-4}							& 3 	& 8.70 		& 1.67 	\\\hline
\hline
\end{tabular}
\caption{{\small Comparison of \Mora{} vs. proof-of-concept (PoC)
    implementation of~\cite{Bartocci2019}.}}
\label{table:compare}
\end{center}
\vspace{-20pt}
\end{wraptable} 

\small
\begin{itemize}\vspace{-5pt}
\item \Mora{} now optimizes the order in which recurrences are 
sent to the {\tt diofant} recurrence solver. This reduces the amount of 
necessary symbolic computation and speeds up the process. 
\item While \Mora{} is implemented entirely in {\tt python}, with limited usage 
of external libraries, the previous implementation was done in {\tt
  Julia} and relied on calls to the {\tt sympy} library of {\tt python}.
\item \Mora{} does not rely on {\tt Aligator}~\cite{Aligator18} for handling systems of 
recurrences, allowing us to  eliminate some intermediate and 
redundant steps.
\end{itemize}\normalsize

\vspace{-7pt}
\section{Conclusion}
\vspace{-5pt}
We described \textsc{Mora}, a fully automated tool for generating
 invariants of probabilistic programs. \Mora{} combines recurrence
 solving,  symbolic
 summation and statistical reasoning, and derives  higher-order moments of  loop variables in probabilistic
 programs.


\vfill

{\small\medskip\noindent{\bf Open Access} This chapter is licensed under the terms of the Creative Commons\break Attribution 4.0 International License (\url{http://creativecommons.org/licenses/by/4.0/}), which permits use, sharing, adaptation, distribution and reproduction in any medium or format, as long as you give appropriate credit to the original author(s) and the source, provide a link to the Creative Commons license and indicate if changes were made.}

{\small \spaceskip .28em plus .1em minus .1em The images or other third party material in this chapter are included in the chapter's Creative Commons license, unless indicated otherwise in a credit line to the material.~If material is not included in the chapter's Creative Commons license and your intended\break use is not permitted by statutory regulation or exceeds the permitted use, you will need to obtain permission directly from the copyright holder.}

\medskip\noindent\includegraphics{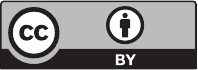}

\end{document}